\documentclass[12pt] {article}
\usepackage{psfrag}
\usepackage{graphicx}
\usepackage{latexsym,amsfonts}
\usepackage{amssymb}
\begin{document}
\setcounter{page}{1}
\def\theequation{\arabic{section}.\arabic{equation}}
\def\theequation{\thesection.\arabic{equation}}
\setcounter{section}{0}

\title{Goldstone bosons in the massless Thirring model.  Witten's
criterion}

\author{M. Faber\thanks{E--mail: faber@kph.tuwien.ac.at, Tel.:
+43--1--58801--14261, Fax: +43--1--58801--14299}~~and
A. N. Ivanov\thanks{E--mail: ivanov@kph.tuwien.ac.at, Tel.:
+43--1--58801--14261, Fax: +43--1--58801--14299}~\thanks{Permanent
Address: State Polytechnical University, Department of Nuclear
Physics, 195251 St. Petersburg, Russian Federation}}

\date{\today}

\maketitle

\vspace{-0.5in}
\begin{center}
{\it Atominstitut der \"Osterreichischen Universit\"aten,
Arbeitsbereich Kernphysik und Nukleare Astrophysik, Technische
Universit\"at Wien, \\ Wiedner Hauptstr. 8-10, A-1040 Wien,
\"Osterreich }
\end{center}

\begin{center}
\begin{abstract}
We discuss the Ward identity and the low--energy theorem for the
divergence of the axial--vector current in the massless Thirring model
with fermion fields quantized in the chirally broken phase
(Eur. Phys. J. C {\bf 20}, 723 (2001)). The Ward identity and the
low--energy theorem are analysed in connection with Witten's criterion
for Goldstone bosons (Nucl. Phys. B {\bf 145}, 110 (1978)). We show
that the free massless (pseudo)scalar field, bosonizing the massless
Thirring model in the chirally broken phase, satisfies Witten's
criterion to interpret quanta of this field as Goldstone bosons. As
has been shown in hep--th/0210104 and hep--th/0212226, Goldstone's
criterion, the non--invariance of the wave function of the ground
state, is also fulfilled.
\end{abstract}
\end{center}

\newpage

\section{Introduction}
\setcounter{equation}{0}

\hspace{0.2in} The massless Thirring model \cite{WT58} is an exactly
solvable quantum field theoretic model of fermions with a non--trivial
four-fermion interaction embedded in 1+1--dimensional space--time. A
solution of a quantum field theoretic model requires the development
of a procedure for the calculation of any correlation function. In the
chiral symmetric phase this has been carried out in \cite{KJ61} (see
also \cite{FI3}).

The existence of the chirally broken phase of the massless Thirring
model has been recently pointed out in \cite{FI1}. It has been shown
that the chirally broken phase of the massless Thirring model is
identical to the superconducting phase of the BCS theory of
superconductivity. In the chirally broken phase the massless Thirring
model bosonizes to the quantum field theory of a free massless
(pseudo)scalar field $\vartheta(x)$. According to
\cite{FI1}--\cite{FI9}, the quanta of the free massless (pseudo)scalar
field $\vartheta(x)$ are Goldstone bosons (see also \cite{IZ80}).

In the paper \cite{EW78} Witten has investigated the $1/N$ expansion
within the massless Thirring model with $SU(N)$ symmetry. He has
confirmed the existence of spontaneously broken chiral symmetry and
massless (pseudo)scalar bosons. However, according to Witten these
bosons are massless but non--Goldstone. Indeed, Witten has written
\cite{EW78}: ``{\it A Goldstone boson is a massless boson whose
singular contributions to the Ward identities enable the identities to
be satisfied even though some symmetry breaking Green functions are
non--zero. This is not possible in two dimensions. But massless bosons
are possible. The massless boson in this theory is not a Goldstone
boson; it satisfies no pertinent low--energy theorems; in this theory
the chirality violating Green functions are zero and there is no room
for Goldstone--boson contributions in the Ward identities}.''

In \cite{AW64} Wightman has pointed out that due to infrared
divergences of the two--point Wightman functions, leading to the
violation of Wightman's positive definiteness condition, one cannot
construct a mathematically correct quantum field theory of a free
massless (pseudo)scalar field in 1+1--dimensional space--time with
Wightman's observables defined on the test functions $h(x)$ from the
Schwartz class ${\cal S}(\mathbb{R}^2)$. In other words, Wightman has
asserted that such a theory does not merely exist. In \cite{SC73}
Coleman has reformulated Wightman's assertion in the form of the
theorem suppressing the existence of Goldstone bosons and 
spontaneous breaking of continuous symmetry in quantum field theories
with Wightman's observables defined on the test functions $h(x)$ from
the Schwartz class ${\cal S}(\mathbb{R}^2)$. This is the well--known
theorem of Coleman \cite{SC73}.

As has been shown in \cite{FI2,FI7}, due to infrared divergences the
{\it vacuum--to--vacuum} transition amplitude (or the generating
functional of Green functions), induced by an external source in
Schwinger's approach to the quantum field theory of the free massless
(pseudo)scalar field, vanishes. This confirms Wightman's statement
\cite{AW64}.

In \cite{FI2} we have shown how one can construct a quantum field
theory of the free massless (pseudo)scalar field in 1+1--dimensional
space--time, which bosonizes the massless Thirring model with fermion
fields quantized in the chirally broken phase, with a non--vanishing
{\it vacuum--to--vacuum} transition amplitude and the two--point
Wightman functions without infrared divergences. As has been shown in
\cite{FI7}, this is equivalent to Wightman's quantum field theory with
Wightman's observables, defined on the test functions from the
Schwartz class ${\cal S}_0(\mathbb{R}^2) = \{h(x) \in {\cal
S}(\mathbb{R}^2); \tilde{h}(0) = 0\}$. Since Wightman's observables
are defined on the test functions from ${\cal S}_0(\mathbb{R}^2)$,
there is no contradiction to Coleman's theorem \cite{FI8,FI9}.

Such a quantum field theory of the free massless (pseudo)scalar field
$\vartheta(x)$ can be applied to the description of massless
(pseudo)scalar bosons in Witten's analysis of the massless Thirring
model.

Now the problem left is: ``Whether one can call the quanta of this
free massless (pseudo)scalar field $\vartheta(x)$ Goldstone bosons or
not ?''  As has been shown in \cite{FI8,FI9}, Goldstone's criterion,
the non--invariance of the wave function of the ground state
\cite{JG61}, is fulfilled. Therefore, in order to gain an allowance to
interpret the quanta of the free massless (pseudo)scalar field
$\vartheta(x)$ as Goldstone bosons, one has to show that they satisfy
Witten's criterion. In order words, one has to demonstrate that the
contributions of the quanta of the $\vartheta$--field play an
important role for the fulfillment of the low--energy theorems and the
Ward identities quoted by Witten.

The paper is organized as follows. In Section 2 we derive the Ward
identity and the low--energy theorem for the divergence of the
axial--vector current. We show that in the chiral symmetric phase,
when the axial--vector current is conserved, the Ward identity and the
low--energy theorem are trivially fulfilled as $0 = 0$. In turn, in
the chirally broken phase the Ward identity and the low--energy
theorem are fulfilled only due to a non--trivial contribution of the
free massless (pseudo)scalar field $\vartheta(x)$.  According to
Witten's criterion this testifies that the quanta of the free massless
(pseudo)scalar field $\vartheta(x)$ are Goldstone bosons. Since the
ground state of a free massless (pseudo)scalar field $\vartheta(x)$
obtained in \cite{FI8} is not invariant under chiral symmetry
transformations, the quanta of the free massless (pseudo)scalar field
$\vartheta(x)$ satisfy also Goldstone's criterion for Goldstone bosons
\cite{JG61}. In Section 3 we show that the Ward identity and the
low--energy theorem derived in Section 2 are consistent with the
gap--equation defining the dynamical mass of the Thirring fermion
fields quantized in the chirally broken phase \cite{FI1}. In the
Conclusion we discuss the obtained results.

\section{Ward's identity and low--energy theorem in the massless 
Thirring model}
\setcounter{equation}{0}

\hspace{0.2in} To clarify the problem whether the quanta of the free
massless (pseudo)scalar field $\vartheta(x)$, the quantum field theory
of which bosonizes the massless Thirring model in the chirally broken
phase \cite{FI1,FI2,FI8}, are Goldstone bosons we suggest to treat the
following vacuum expectation value
\begin{eqnarray}\label{label2.1}
\langle 0|{\rm T}(j^{\mu}_5(x)\psi(y)\bar{\psi}(z))|0\rangle_{\rm
conn.},
\end{eqnarray}
where $j^{\mu}_5(x) = \bar{\psi}(x)\gamma^{\mu}\gamma^5\psi(x)$ is the
axial--vector current, $\psi(x)$ and $\bar{\psi}(x)$ are massless
Thirring fermion fields\footnote{The Dirac $\gamma$--matrices are
defined by \cite{FI1}: $\gamma^{\mu} = (\gamma^0 = \sigma_1, \gamma^1
= -i\sigma_2)$ and $\gamma^5 = \gamma^0\gamma^1 = \sigma_3$, where
$\sigma_i\,(i = 1,2,3)$ are Pauli matrices .}. For the derivation of
the Ward identity we need to write down explicitly the time--ordered
product. It reads
\begin{eqnarray}\label{label2.2}
{\rm T}(j^{\mu}_5(x)\psi_a(y)\bar{\psi}_b(z)) &=& \theta(x^0 -
y^0)\theta(y^0 - z^0)\, j^{\mu}_5(x) \psi_a(y)
\bar{\psi}_b(z)\nonumber\\ &-& \theta(z^0 - x^0)\theta(x^0 - y^0)\,
\bar{\psi_b}(z) j^{\mu}_5(x) \psi_a(y) \nonumber\\ &+& \theta(y^0 -
z^0)\theta(z^0 - x^0)\, \psi_a(y) \bar{\psi}_b(z)j
^{\mu}_5(x)\nonumber\\ &+& \theta(y^0 - x^0)\theta(x^0 -
z^0)\,\psi_a(y) j^{\mu}_5(x) \bar{\psi}_b(z) \nonumber\\ &-&
\theta(x^0 - z^0)\theta(z^0 - y^0)\, j^{\mu}_5(x) \bar{\psi}_b(z)
\psi_a(y)\nonumber\\ &-& \theta(z^0 - y^0)\theta(y^0 - x^0)\,
\bar{\psi}_b(z)\psi_a(y) j^{\mu}_5(x)
\end{eqnarray}
The divergence of this time--ordered product is equal to
\begin{eqnarray}\label{label2.3}
\partial_{\mu}{\rm T}(j^{\mu}_5(x)\psi_a(y)\bar{\psi}_b(z)) &=& {\rm
T}(\partial_{\mu}j^{\mu}_5(x)\psi_a(y)\bar{\psi}_b(z)) + \delta(x^0 -
y^0)\,{\rm T}([j^0_5(x), \psi_a(y)] \bar{\psi}_b(z))\nonumber\\ &&+
\delta(x^0 - z^0)\,{\rm T}(\psi_a(y)[j^0_5(x), \bar{\psi}_b(z)]).
\end{eqnarray}
This relation is valid for Thirring fermion fields quantized in both
chiral symmetric and chirally broken phases.

The equal--time commutation relations can be calculated using the
canonical anti--commutation relations
\begin{eqnarray}\label{label2.4}
\delta(x^0 - y^0)\,\{\psi_a(x),\psi^{\dagger}_b(y)\} =
\delta_{ab}\,\delta^{(2)}(x - y).
\end{eqnarray}
This gives
\begin{eqnarray}\label{label2.5}
\delta(x^0 - y^0)[j^0_5(x), \psi_a(y)] &=& - \delta^{(2)}(x -
y)\,\gamma^5_{ac}\psi_c(y)\nonumber\\ \delta(x^0 - z^0)[j^0_5(x),
\bar{\psi}_b(z)] &=&- \delta^{(2)}(x - z)\,\bar{\psi}_c(z)
\gamma^5_{cb},
\end{eqnarray}
where we have used the relation $[AB,C] = A\{B,C\} - \{A,C\}B$.

Substituting (\ref{label2.5}) in (\ref{label2.3}) we get
\begin{eqnarray}\label{label2.6}
&&\partial_{\mu}{\rm T}(j^{\mu}_5(x)\psi_a(y)\bar{\psi}_b(z)) = {\rm
T}(\partial_{\mu}j^{\mu}_5(x)\psi_a(y)\bar{\psi}_b(z))\nonumber\\ && -
\delta^{(2)}(x - y)\,{\rm T}(\gamma^5_{ac}\psi_c(y)\bar{\psi}_b(z)) -
\delta^{(2)}(x - z)\,{\rm T}(\psi_a(y)\bar{\psi}_c(z)\gamma^5_{cb}).
\end{eqnarray}
The corresponding Ward identity reads
\begin{eqnarray}\label{label2.7}
\hspace{-0.5in}&&\partial_{\mu}\langle 0|{\rm
T}(j^{\mu}_5(x)\psi_a(y)\bar{\psi}_b(z))|0\rangle_{\rm conn.} =
\langle 0|{\rm
T}(\partial_{\mu}j^{\mu}_5(x)\psi_a(y)\bar{\psi}_b(z))|0\rangle_{\rm
conn.}\nonumber\\ \hspace{-0.5in}&&- \delta^{(2)}(x - y)\,\langle
0|{\rm T}(\gamma^5_{ac}\psi_c(y)\bar{\psi}_b(z))|0\rangle -
\delta^{(2)}(x - z)\,\langle 0|{\rm
T}(\psi_a(y)\bar{\psi}_c(z)\gamma^5_{cb})|0\rangle.
\end{eqnarray}
Integrating both sides over $x$ we obtain
\begin{eqnarray}\label{label2.8}
&&\int d^2x\,\partial_{\mu}\langle 0|{\rm
T}(j^{\mu}_5(x)\psi_a(y)\bar{\psi}_b(z))|0\rangle_{\rm conn.} = \int
d^2x\,\langle 0|{\rm T}(\partial_{\mu} j^{\mu}_5(x) \psi_a(y)
\bar{\psi}_b(z)) |0\rangle_{\rm conn.} \nonumber\\ &&- \langle 0|{\rm
T}(\gamma^5_{ac}\psi_c(y)\bar{\psi}_b(z))|0\rangle - \langle 0|{\rm
T}(\psi_a(y)\bar{\psi}_c(z)\gamma^5_{cb})|0\rangle.
\end{eqnarray}
Assuming that the surface term does not give a contribution, the Ward
identity (\ref{label2.8}) can be reduced to the form
\begin{eqnarray}\label{label2.9}
\int d^2x\,\langle 0|{\rm
T}(\partial_{\mu}j^{\mu}_5(x)\psi_a(y)\bar{\psi}_b(z))|0\rangle_{\rm
conn.} &=& \langle 0|{\rm
T}(\gamma^5_{ac}\psi_c(y)\bar{\psi}_b(z))|0\rangle\nonumber\\ &+&
\langle 0|{\rm T}(\psi_a(y)\bar{\psi}_c(z)\gamma^5_{cb})|0\rangle.
\end{eqnarray}
For the subsequent analysis it is convenient to set $z = 0$. This
yields
\begin{eqnarray}\label{label2.10}
\int d^2x\,\langle 0|{\rm
T}(\partial_{\mu}j^{\mu}_5(x)\psi_a(y)\bar{\psi}_b(0))|0\rangle_{\rm
conn.} &=& \langle 0|{\rm
T}(\gamma^5_{ac}\psi_c(y)\bar{\psi}_b(0))|0\rangle\nonumber\\ &+&
\langle 0|{\rm T}(\psi_a(y)\bar{\psi}_c(0)\gamma^5_{cb})|0\rangle.
\end{eqnarray}
The vacuum expectation values in the r.h.s. of (\ref{label2.10}) are
related to the two--point fermion Green function, which we define using
the K\"allen--Lehmann representation \cite{KL}
\begin{eqnarray}\label{label2.11}
\langle 0|{\rm T}(\psi_a(y)\bar{\psi}_b(0))|0\rangle = \int
\frac{d^2k}{(2\pi)^2i}\,e^{\textstyle\,-ik\cdot
y}\int^{\infty}_0dm^2\,\frac{\hat{k}_{ab}\,\rho_1(m^2) +
\delta_{ab}\,\rho_2(m^2)}{m^2 - k^2 - i\,0},
\end{eqnarray}
where $\rho_1(m^2)$ and $\rho_2(m^2)$ are K\"allen--Lehmann spectral
functions. It is obvious that in the chiral symmetric phase
$\rho_2(m^2) = 0$.

In terms of the K\"allen--Lehmann representation the r.h.s. of
(\ref{label2.10}) can be written as
\begin{eqnarray}\label{label2.12}
\hspace{-0.3in}&&\langle 0|{\rm
T}(\gamma^5_{ac}\psi_c(y)\bar{\psi}_b(0))|0\rangle + \langle 0|{\rm
T}(\psi_a(y)\bar{\psi}_c(0)\gamma^5_{cb})|0\rangle =\nonumber\\
\hspace{-0.3in}&&= \int
\frac{d^2k}{(2\pi)^2i}\,e^{\textstyle\,-ik\cdot
y}\int^{\infty}_0\frac{dm^2}{m^2 - k^2 - i\,0}\,\{(\gamma^5\hat{k} +
\hat{k}\gamma^5)_{ab}\,\rho_1(m^2) + 2\,\gamma^5_{ab}\rho_2(m^2)\} =
\nonumber\\
\hspace{-0.3in}&&= 2 \gamma^5_{ab}\int
\frac{d^2k}{(2\pi)^2i}\,e^{\textstyle\,-ik\cdot
y}\int^{\infty}_0dm^2\,\frac{\rho_2(m^2)}{m^2 - k^2 - i\,0}.
\end{eqnarray}
Substituting (\ref{label2.12}) in (\ref{label2.10}), multiplying by
$e^{\textstyle\,ip\cdot y}$ and integrating over $y$ we get
\begin{eqnarray}\label{label2.13}
\hspace{-0.3in}i\int\!\!\!\int d^2x\,d^2y\,e^{\textstyle\,ip\cdot
y}\,\langle 0|{\rm
T}(\partial_{\mu}j^{\mu}_5(x)\psi_a(y)\bar{\psi}_b(0))|0\rangle_{\rm
c} = 2\gamma^5_{ab}\int^{\infty}_0dm^2\,\frac{\rho_2(m^2)}{m^2 -
p^2 - i\,0}.
\end{eqnarray}
Now we suggest to multiply both sides of (\ref{label2.13}) by
$\gamma^5_{ba}$, to sum over $a$ and $b$ and to take the low--energy
limit $ p \to 0$.  This gives the following low--energy theorem
\begin{eqnarray}\label{label2.14}
\frac{i}{4}\int\!\!\!\int d^2x\,d^2y\,\langle 0|{\rm
T}(\partial_{\mu}j^{\mu}_5(x)\psi_a(y)\bar{\psi}_b(0)\gamma^5_{ba})
|0\rangle_{\rm conn.} = \int^{\infty}_0\frac{dm^2}{m^2}\,\rho_2(m^2).
\end{eqnarray}
In the chiral symmetric phase when the axial--vector current is
conserved, $\partial_{\mu}j^{\mu}_5(x) = 0$, and $\rho_2(m^2) = 0$ the
low--energy theorem (\ref{label2.14}) reduces to a trivial equality $0
= 0$.

Now let us consider the fulfillment of the low--energy theorem in the
chirally broken phase. According to \cite{FI1}, in the chirally broken
phase the Thirring fermion fields obey the equations of motion
\begin{eqnarray}\label{label2.15}
\gamma^{\mu}\partial_{\mu}\psi(x) &=& - i
M\,e^{\textstyle\,i\gamma^5\beta \vartheta(x)}\,\psi(x),\nonumber\\
\partial_{\mu}\bar{\psi}(x)\gamma^{\mu} &=& + iM\,\bar{\psi}(x)\,
e^{\textstyle\,i\gamma^5 \beta \vartheta(x)},
\end{eqnarray}
where $M$ is a dynamical mass of the massless Thirring fermion fields
quantized relative to the non--perturbative vacuum \cite{FI1}. The
coupling constant $\beta$ is related to the coupling constant $g$ of
the massless Thirring model as \cite{FI1,FI5}
\begin{eqnarray}\label{label2.16}
\frac{8\pi}{\beta^2} = 1 - e^{\textstyle\,-2\pi/g}.
\end{eqnarray}
The divergence of the axial--vector current is equal to
\begin{eqnarray}\label{label2.17}
\partial_{\mu}j^{\mu}_5(x) = 2M\,\bar{\psi}(x)i\gamma^5
e^{\textstyle\,i\gamma^5\beta\vartheta(x)}\psi(x).
\end{eqnarray}
The vacuum expectation value in the l.h.s. of
(\ref{label2.14}) we represent in terms of the generating functional
of Green functions of the massless Thirring fermion fields \cite{FI1}
\begin{eqnarray}\label{label2.18}
\hspace{-0.3in}&&-\,M\,\frac{1}{2}\int\!\!\!\int d^2xd^2y
\Big(\gamma^5\,\exp\Big\{\gamma^5\beta\frac{\delta}{\delta
J(x)}\Big\}\Big)_{cd}\gamma^5_{ba}\frac{\delta}{\delta
\eta_c(x)}\frac{\delta}{\delta \bar{\eta}_d(x)}\frac{\delta}{\delta
\bar{\eta}_a(y)}\frac{\delta}{\delta \eta_b(0)}\nonumber\\
\hspace{-0.3in}&&\times\,Z[\eta,\bar{\eta},J]\Big|_{\eta = \bar{\eta} =
J = 0} = \int^{\infty}_0\frac{dm^2}{m^2} \rho_2(m^2),
\end{eqnarray}
where $Z[\eta,\bar{\eta},J]$ is the generating functional of Green
functions defined by \cite{FI1,FI2}
\begin{eqnarray}\label{label2.19}
\hspace{-0.3in}Z[\eta,\bar{\eta},J] &=& \int {\cal
D}\vartheta\,\exp\Big\{i\int d^2z \Big[\frac{1}{2}
\partial_{\nu}\vartheta(z)\partial^{\nu}\vartheta(z) +
\vartheta(z)\,J(z)\,\Big]\nonumber\\ &&+ i\int\!\!\!\int
d^2z_1d^2z_2\,\bar{\eta}(z_1)\,S_F(z_1,z_2;\vartheta)\,\eta(z_2)\Big\}.
\end{eqnarray}
Here $S_F(z_1,z_2;\vartheta)$ is the fermion Green function obeying the
equation 
\begin{eqnarray}\label{label2.20}
\Big(i\gamma^{\alpha}\frac{\partial}{\partial z^{\alpha}_1} -
M\,e^{\textstyle\,i\gamma^5\beta\vartheta(z_1)}\Big)
S_F(z_1,z_2;\vartheta) = - \delta^{(2)}(z_1 - z_2),
\end{eqnarray}
and $J(x)$ is an external source of the $\vartheta$--field
satisfying the constraint \cite{FI2}
\begin{eqnarray}\label{label2.21}
\int d^2x\,J(x) = \tilde{J}(0) = 0.
\end{eqnarray}
Here $\tilde{J}(0)$ is the Fourier transform of the external source
$J(x)$ at momentum zero, $k = 0$. Due to the constraint
(\ref{label2.21}) the collective zero--mode, describing the motion of
the ``center of mass'' of the free massless (pseudo)scalar field
$\vartheta(x)$, does not contribute to correlation functions
\cite{FI2,FI8}. This is important, since, as has been shown in
\cite{FI9}, the collective zero--mode is responsible for the infrared
divergences of the two--point Wightman functions in the quantum field
theory of the free massless (pseudo)scalar field $\vartheta(x)$.

Differentiating in (\ref{label2.18}) with respect to the external
sources of fermion fields we obtain
\begin{eqnarray}\label{label2.22}
\hspace{-0.3in}M\,\frac{1}{2}\int\!\!\!\int d^2xd^2y \Big\langle{\rm
tr}\{S_F(y,x;\vartheta)\gamma^5
\,e^{\textstyle\,i\gamma^5\beta\vartheta(x)}S_F(x, 0; \vartheta)
\gamma^5\}\Big\rangle_{\rm conn.} = \int^{\infty}_0\frac{dm^2}{m^2}
\rho_2(m^2).
\end{eqnarray}
Since the dynamical mass $M$ is proportional to the ultra--violet
cut--off $\Lambda$ \cite{FI1}, we can neglect the contribution of the
gradient term and use the Green function equal to
\begin{eqnarray}\label{label2.23}
S_F(y,x;\vartheta) = \frac{1}{M}\,
e^{\textstyle\,-i\gamma^5\beta\vartheta(y)}\, \delta^{(2)}(y - x).
\end{eqnarray}
Substituting (\ref{label2.23}) in (\ref{label2.22}) and calculating
the trace over Dirac matrices we arrive at the low--energy theorem
\begin{eqnarray}\label{label2.24}
\frac{1}{M}\,\langle \cos\beta\vartheta(0)\rangle =
\int^{\infty}_0\frac{dm^2}{m^2} \rho_2(m^2).
\end{eqnarray}
Recall that the vacuum expectation value $\langle
\cos\beta\vartheta(0)\rangle$ has the meaning of a spontaneous
magnetization \cite{FI2}. It is related to the fermion condensate
$\langle \bar{\psi}\psi\rangle$ by \cite{FI1,FI2}.
\begin{eqnarray}\label{label2.25}
\langle \bar{\psi}\psi\rangle = - \frac{M}{g}\,\langle
\cos\beta\vartheta(0)\rangle,
\end{eqnarray}
Thus, the low--energy theorem (\ref{label2.24}) defines the fermion
condensate as
\begin{eqnarray}\label{label2.26}
\langle \bar{\psi}\psi\rangle = -
\frac{\,M^2}{g}\int^{\infty}_0\frac{dm^2}{m^2}\,\rho_2(m^2).
\end{eqnarray}
As has been shown in \cite{FI2}, $\langle \bar{\psi}\psi\rangle = -
M/g$, therefore we get
\begin{eqnarray}\label{label2.27}
\frac{1}{M} = \int^{\infty}_0\frac{dm^2}{m^2}\,\rho_2(m^2).
\end{eqnarray}
This testifies that the quanta of the free massless (pseudo)scalar
field $\vartheta(x)$, describing the bosonized form of the massless
Thirring model in the chirally broken phase, behave like Goldstone
bosons and saturate the low--energy theorem defining the spontaneous
magnetization and the fermion condensate.

In order to argue that the quanta of the massless (pseudo)scalar field
$\vartheta(x)$ satisfy Witten's criterion for Goldstone bosons
\cite{EW78} we have to show that the low--energy theorem
(\ref{label2.27}) agrees with the gap--equation defining the dynamical
mass of the Thirring fermion fields in the chirally broken phase
\cite{FI1}.

\section{Gap--equation and  low--energy theorem}
\setcounter{equation}{0}

\hspace{0.2in} As has been shown in \cite{FI1}, the dynamical mass of
the Thirring fermion fields quantized in the chirally broken phase
obeys the gap--equation
\begin{eqnarray}\label{label3.1}
M = M\,\frac{g}{2\pi}\,{\ell n}\Big(1 + \frac{\Lambda^2}{M^2}\Big),
\end{eqnarray}
where $\Lambda$ is an ultra--violet cut--off.

A consistency of the low--energy theorem (\ref{label2.27}) with the
gap--equation should confirm the interpretation of the quanta of the
free massless (pseudo)scalar field $\vartheta(x)$ as Goldstone bosons.

In order to proof the agreement of the low--energy theorem
(\ref{label2.27}) and the gap--equation (\ref{label3.1}) we suggest to
turn to the analysis of the spontaneously broken chiral symmetry in
the massless Thirring model in terms of the normal ordering
\cite{FI1}. The Lagrangian of the massless Thirring model reads
\cite{FI1}
\begin{eqnarray}\label{label3.2}
{\cal L}(x)
=:\bar{\psi}(x)i\gamma^{\nu}\partial_{\nu}\psi(x):_{\mu} -
\frac{1}{2}\,g :\bar{\psi}(x)\gamma_{\nu}\psi(x)
\bar{\psi}(x)\gamma^{\nu}\psi(x):_{\mu},
\end{eqnarray}
where $:\ldots:$ denotes the normal ordering at the infrared scale
$\mu \to 0$. The Lagrangian (\ref{label3.2}) can be also rewritten as
\cite{FI1}
\begin{eqnarray}\label{label3.3}
{\cal L}(x)
=:\bar{\psi}(x)i\gamma^{\nu}\partial_{\nu}\psi(x):_{\mu} -
\frac{1}{2}\,g :\bar{\psi}(x)\gamma_{\nu}\psi(x):_{\mu}:
\bar{\psi}(x)\gamma^{\nu}\psi(x):_{\mu}.
\end{eqnarray}
Changing the scale of the normal ordering from $\mu \to 0$ to $M$ we
get
\begin{eqnarray}\label{label3.4}
{\cal L}(x) &=&:\bar{\psi}(x)(i\gamma^{\nu}\partial_{\nu} -
M)\psi(x):_M + (M -
g\gamma_{\nu}(-i)S_F(0)\gamma^{\nu}):\bar{\psi}(x)\psi(x):_M
\nonumber\\ && - \frac{1}{2}\,g :\bar{\psi}(x)\gamma_{\nu}\psi(x)
\bar{\psi}(x)\gamma^{\nu}\psi(x):_M,
\end{eqnarray}
where the normal ordering should be carried out at the scale $M$;
$S_F(0)$ is the total two--point Green function, which we define in the
K\"allen--Lehmann representation (\ref{label2.11}). This yields
\begin{eqnarray}\label{label3.5}
\gamma_{\nu}(-i)S_F(0)\gamma^{\nu} &=&
\int^{\infty}_0dm^2\,\rho_2(m^2)\int \frac{d^2k}{2\pi^2
i}\,\frac{1}{m^2 - k^2 - i\,0} =\nonumber\\ &=&
\frac{1}{2\pi}\int^{\infty}_0dm^2\,\rho_2(m^2)\,{\ell n}\Big(1 +
\frac{\Lambda^2}{m^2}\Big).
\end{eqnarray}
Substituting (\ref{label3.5}) in (\ref{label3.4}) we arrive at
\begin{eqnarray}\label{label3.6}
{\cal L}(x) &=&:\bar{\psi}(x)(i\gamma^{\nu}\partial_{\nu} -
M)\psi(x):_M\nonumber\\ &&+ \Big[M -
\frac{g}{2\pi}\int^{\infty}_0dm^2\,\rho_2(m^2)\,{\ell n}\Big(1 +
\frac{\Lambda^2}{m^2}\Big)\Big]:\bar{\psi}(x)\psi(x):_M \nonumber\\ &&
- \frac{1}{2}\,g :\bar{\psi}(x)\gamma_{\nu}\psi(x)
\bar{\psi}(x)\gamma^{\nu}\psi(x):_M.
\end{eqnarray}
The scale $M$ acquires the meaning of a dynamical mass if it satisfies
the gap--equation
\begin{eqnarray}\label{label3.7}
M - \frac{g}{2\pi}\int^{\infty}_0dm^2\,\rho_2(m^2)\,{\ell n}\Big(1 +
\frac{\Lambda^2}{m^2}\Big) = 0.
\end{eqnarray}
Due to the gap--equation (\ref{label3.7}) the Lagrangian
(\ref{label3.6}) takes the form
\begin{eqnarray}\label{label3.8}
{\cal L}(x) = :\bar{\psi}(x)(i\gamma^{\nu}\partial_{\nu} -
M)\psi(x):_M - \frac{1}{2}\,g :\bar{\psi}(x)\gamma_{\nu}\psi(x)
\bar{\psi}(x)\gamma^{\nu}\psi(x):_M,
\end{eqnarray}
where $M$ can be identified with the mass of the fermion fields.

The gap--equation (\ref{label3.7}) and the low--energy theorem
(\ref{label2.27}) are consistent if the K\"allen--Lehmann spectral
function $\rho_2(m^2)$ is equal to
\begin{eqnarray}\label{label3.9}
\rho_2(m^2) = M\,\delta(m^2 - M^2).
\end{eqnarray}
In the chiral symmetric phase when $M = 0$ the spectral function
$\rho_2(m^2)$ vanishes.

This completes the proof of the agreement of the low--energy theorem
(\ref{label2.27}) (or more general (\ref{label2.14})) and the
gap--equation (\ref{label3.1}) for the dynamical mass of the Thirring
fermion fields quantized in the chirally broken phase.

Thus, we can argue that the quanta of the free massless (pseudo)scalar
field $\vartheta(x)$, bosonizing the massless Thirring model in the
chirally broken phase \cite{FI1,FI2} and described by the quantum
field theory without infrared divergences \cite{FI2}--\cite{FI9},
satisfy Witten's criterion for Goldstone bosons.

\section{Conclusion}

\hspace{0.2in} The quantum field theory of the free massless
(pseudo)scalar field $\vartheta(x)$, described by the Lagrangian
${\cal L}(x) = \frac{1}{2}\, \partial_{\mu}\vartheta(x)
\partial^{\mu}\vartheta(x)$, bosonizes the massless Thirring model
with fermion fields quantized in the chirally broken phase. The main
aim of this paper is to show that the quanta of this field satisfy
Witten's criterion for Goldstone bosons \cite{EW78}. According to
Witten \cite{EW78}, the quanta of the massless (pseudo)scalar field
$\vartheta(x)$ should saturate the low--energy theorems.  Following
Witten's requirement we have derived a low--energy theorem for the
divergence of the axial--vector current. We have shown that in the
chiral symmetric phase of the massless Thirring model, when the
divergence of the axial--vector current vanishes, this low--energy
theorem reduces to a trivial identity $0 = 0$. In turn, in the
chirally broken phase with a non--vanishing divergence of the
axial--vector current, the fulfillment of the low--energy theorem is
fully caused by the contribution of the free massless (pseudo)scalar
field $\vartheta(x)$ in terms of the spontaneous magnetization
$\langle \cos\beta\vartheta(0)\rangle$. In the quantum field theory of
the free massless (pseudo)scalar field $\vartheta(x)$ without infrared
divergences this magnetization does not vanish, $\langle
\cos\beta\vartheta(0)\rangle = 1$ \cite{FI2,FI8}. This defines a
non--vanishing fermion condensate in the massless Thirring model with
fermion fields quantized in the chirally broken phase, $\langle
\bar{\psi}\psi\rangle = - \langle\cos\beta\vartheta(0)\rangle\,M/g = -
M/g$ \cite{FI1,FI2}. 

The consistency of the obtained results with the existence of the
chirally broken phase of the massless Thirring model has been
confirmed by the agreement of the low--energy theorem under
consideration with the gap--equation, defining a dynamical mass of the
Thirring fermion fields quantized in the chirally broken phase.
Hence, one can conclude that the quanta of the free massless
(pseudo)scalar field $\vartheta(x)$ satisfy Witten's criterion for
Goldstone bosons.

As has been shown in \cite{FI8,FI9}, the wave function of the ground
state of the free massless (pseudo)scalar field $\vartheta(x)$ is
described by the bosonized BCS--type wave function of the ground state
of the massless Thirring model in the chirally broken phase. This wave
function is not invariant under chiral symmetry transformations
\cite{FI8,FI9}. Such a non--invariance agrees with Goldstone's
criterion \cite{JG61} allowing to interpret the quanta of the
$\vartheta$--field as Goldstone bosons.

Thus, the quanta of the free massless (pseudo)scalar field
$\vartheta(x)$, bosonizing the massless Thirring model with fermion
fields quantized in the chirally broken phase \cite{FI1}--\cite{FI9},
satisfy both Witten's and Goldstone's criteria for Goldstone bosons.

We would like to emphasize that this is not a counterexample to
Coleman's theorem \cite{SC73}. Indeed, the quantum field theory of the
free massless (pseudo)scalar field $\vartheta(x)$ in 1+1--dimensional
space--time is equivalent to Wightman's quantum field theory with
Wightman's observables defined on the test functions $h(x)$ from the
Schwartz class ${\cal S}_0(\mathbb{R}^2) = \{h(x) \in {\cal
S}(\mathbb{R}^2); \tilde{h}(0) = 0\}$ \cite{FI7}--\cite{FI9}, whereas
Coleman's theorem \cite{SC73} has been formulated for the quantum
field theories in 1+1--dimensional space--time with Wightman's
observables defined on the test functions from the Schwartz class
${\cal S}(\mathbb{R}^2)$.


\begin{thebibliography}{9}
\bibitem{WT58} 
W. Thirring, 
Ann. Phys. (N.Y.) {\bf 3}, 91 (1958).
\bibitem{KJ61}
K. Johnson,
Nuovo Cim. {\bf 20}, 773 (1961);
F. L. Scarf and J. Wess,
Nuovo Cim. {\bf 26}, 150 (1962);
C. R. Hagen, 
Nuovo Cim. {\bf 51}, 169 (1967);
B. Klaiber, 
in {\it LECTURES IN THEORETICAL PHYSICS},
Lectures delivered at the Summer Institute for Theoretical Physics,
University of Colorado, Boulder, 1967, edited by A. Barut and
W. Brittin, Gordon and Breach, New York, 1968, Vol. X, 
part A, pp.141--176.
\bibitem{FI3}
M. Faber and A. N. Ivanov,
{\it On the solution of the massless
Thirring model  with fermion
fields quantized in the chiral symmetric phase}, hep--th/0112183.
\bibitem{FI1}
M. Faber and A. N. Ivanov, 
Eur. Phys. J. C {\bf 20}, 723 (2001).
\bibitem{FI2}
M. Faber and A. N. Ivanov, 
Eur. Phys. J. C {\bf 24}, 653 (2002).
\bibitem{FI4}
M. Faber and A. N. Ivanov,
{\it Comments no Coleman's paper ``There are no Goldstone Bosons in
Two Dimensions''},
hep--th/0204237.
\bibitem{FI5}
M. Faber and A. N. Ivanov, 
{\it Is the energy of the ground state of the sine--Gordon model 
unbounded from below for $\beta^2 > 8\pi$} ?, hep--th/0205249,
2002.
\bibitem{FI6} 
M. Faber and A. N. Ivanov,
{\it Massless Thirring fermion fields in the boson field 
representation}, hep--th/0206034, 2002.
\bibitem{FI7} 
M. Faber and A. N. Ivanov,
{\it Quantum field theory of a free massless (pseudo)scalar field 
in 1+1--dimensional space--time as a test for the massless Thirring 
model}, hep--th/0206244, 2002.
\bibitem{FI8}
M. Faber and A. N. Ivanov, 
{\it Bosonic vacuum wave functions from the BCS--type wave function
of the ground state of the massless Thirring model}, hep--th/0210104,
(to appear in Physics Letters B).
\bibitem{FI9} 
M. Faber and A. N. Ivanov,
{\it On the ground state of the massless (pseudo)scalar field in two
dimensions}, hep--th/0212226, 2002.
\bibitem{IZ80}
C. Itzykson and J.--B. Zuber,
in {\it QUANTUM FIELD THEORY}, McGraw--Hill Book Company, 
New York, 1980, p.521.
\bibitem{EW78}
E. Witten,
Nucl. Phys. B {\bf 145}, 110 (1978).
\bibitem{AW64} 
A. S. Wightman, 
{\it Introduction to Some Aspects of
the Relativistic Dynamics of Quantized Fields}, in {\it HIGH ENERGY
ELECTROMAGNETIC INTERACTIONS AND FIELD THEORY}, Carg$\grave{\rm e}$se
Lectures in Theoretical Physics, edited by M. Levy , 1964, Gordon and
Breach, 1967, pp.171--291;
R. F. Streater and A. S. Wightman,
in {\it PCT, SPIN AND STATISTICS, AND ALL THAT},
Princeton University Press, Princeton and Oxford, Third Edition, 
1980.
\bibitem{SC73}
S. Coleman,
Comm. Math. Phys. {\bf 31}, 259 (1973).
\bibitem{JG61}
J. Goldstone,
Nuovo Cimento {\bf 19}, 154 (1961);
J. Goldstone, A. Salam, and S. Weinberg,
Phys. Rev. {\bf 127}, 965 (1962);
(see \cite{IZ80} pp.519--526).
\bibitem{KL}
(see \cite{IZ80} p.214).
\end{thebibliography}
\end{document}